\documentclass[a4paper,fleqn,usenatbib]{mnras}

\usepackage{newtxtext,newtxmath}

\usepackage[T1]{fontenc}
\usepackage{ae,aecompl}

\usepackage{siunitx}
\usepackage{graphicx}	
\usepackage{amsmath}	
\usepackage{amssymb}	

\title{Superdiffusive  Stochastic Fermi Acceleration in Space and Energy}

\author[N. Sioulas et al.]{
N. Sioulas, H. Isliker, L. Vlahos, A. Koumtzis and Th. Pisokas\\
Department of Physics,  Aristotle University,
GR-52124 Thessaloniki, Greece
}

\date{Accepted XXX. Received YYY; in original form ZZZ}

\pubyear{2019}

\begin{document}
\label{firstpage}
\pagerange{\pageref{firstpage}--\pageref{lastpage}}
\maketitle

\begin{abstract}
We analyze the transport properties of charged  particles (ions and electrons) interacting with randomly formed magnetic scatterers (e.g.\ large scale local ``magnetic fluctuations'' or ``coherent magnetic irregularities'' usually present in strongly turbulent plasmas), using the energization processes proposed initially by  Fermi in 1949. The scatterers are formed  by  large scale local fluctuations  ($\delta B/B \approx 1$) and are randomly distributed inside the unstable magnetic topology. We construct a 3D grid on which a small fraction of randomly chosen grid points are acting as scatterers. In particular, we study how a large number of test particles are accelerated and transported inside a  collection of scatterers  in a finite volume.  Our main results are: (1) The spatial mean-square  displacement $<(\Delta r)^2>$ inside the stochastic Fermi accelerator is superdiffusive,   $<(\Delta r)^2> \sim t^{a_{r}},$ with $a_r \sim 1.2-1.6$, for the high energy electrons with kinetic energy $(W)$ larger than $1 MeV$,  and it is normal ($a_r=1$) for  the heated low energy $(W< 10 keV)$ electrons. (2) The transport properties of the high energy particles are closely related with the mean-free path that the particles travel in-between the scatterers  ($\lambda_{sc}$). The smaller $\lambda_{sc}$ is, the faster the electrons and ions escape from the acceleration volume.  (3) 
	The mean  displacement in energy $<\Delta W> \sim t^{a_{W}}$ is strongly enhanced inside the acceleration volume $(a_W=1.5- 2.5)$ for the high energy particles compared to the thermal  low energy particles ($a_W=0.4$), i.e.\ high energy particles undergo an enhanced systematic gain in energy.
(4) The mean-square displacement in energy $<W^2>$ is superdiffusive for the high energy particles and normal for the low energy, heated particles.  (5) The acceleration time is also related with  $\lambda_{sc}$. (6) Electrons and ions have similar transport  properties inside the stochastic Fermi accelerator. 

\end{abstract}



\section{Introduction} 
Stochastic Fermi acceleration is a very broad term for particles  interacting with any kind of scattering centers, e.g.\ large amplitude MHD disturbances, current sheets, and waves from electromagnetic (em) resonances.  The interaction of particles with waves was called ``the modern version'' of stochastic Fermi acceleration in order to distinguish it from the original version proposed by Fermi in 1949 \citep{Fermi49}.

The interaction of particles with em fields  inside an unstable magnetized plasma is an important process for the acceleration and transport of particles in laboratory and astrophysical plasmas. In realistic space, astrophysical, or laboratory plasmas, the 3D evolution of the em fields is usually  analyzed, up to now, with the use of the resistive MHD equations. The evolution of the particles inside the em fields is followed by using either the test particle approach or with the use of transport equations. 

Depending on the amplitude of the fluctuating em fields, the interaction of the particles and fields is dramatically different.
For small amplitude  plasma waves the dispersion relation for the linearized system ($\omega=\omega(\vec{k})$) is valid and the magnetic fluctuations can be expressed as 
\begin{equation}\label{spectrum}
	\delta \vec{B}(\vec{x},t) =\sum_{k_{min}}^{k_{max}} \vec{B}_{k0} \;  exp[i (\vec{k}\cdot \vec{x}-\omega(\vec{k})t+ \phi_{k}] ,
	\end{equation}
where $\vec B_{k0}$ is the amplitude of the waves, $\vec k$ is the wave vector and $\phi_k$ is the random phase of the specific wave mode.
The magnetic field is given through the expression $\vec{B}(\vec{x},t)=\vec{B}_0+\delta \vec{B}(\vec{x},t)$ and the electric field writes $\vec{E}(\vec{x},t)=0+\delta \vec{E}(\vec{x},t)$, where
$$ \mathrm{rot} \; \delta \vec{E}=-c^{-1}  \frac{\partial \delta \vec{B}}{\partial t},$$
 $\vec{B}_0$ is the ambient magnetic field, and the ensemble averages are $<\delta \vec{B}>=<\delta \vec{E}> = 0.$   Several analytical and numerical  tools are available for the study the wave-particle interaction when the (possibly unstable) em fluctuations are weak ($\vert \delta \vec{B} \vert << \vert \vec{B}_0 \vert. $)  We must stress here that only weak turbulence can be represented by a discrete number of  normal modes of the plasma. 

Stochastic (second order) Fermi acceleration of particles can be analyzed with at least two very well known techniques: (1)  {\bf Test particle} simulations  inside em fluctuations described by Eq.\ \ref{spectrum} (\citet{Perri2011, Perri09, Greco10}).
(2) The solution of the  {\bf quasilinear} (QL) equation  \citep{Kennel66}. The QL equation is used extensively for the analysis of the interaction of charged particles with partially random em fields. The QL equation is derived from the first order expansion of the fluctuating em fields $(\delta B/B_0)^2,$ assuming that all higher order terms are zero in weak turbulence. The limits of the QL approximation have been explored by using test particle simulations \citep{Kuramitsu00, Zacharegkas16}. The standard QL approach also assumes incoherent mode coupling of the fluctuating em fields, described by the superposition of individual plasma wave modes. The em fields are assumed  known  and not evolving, so the response of the particle energy distribution can be calculated (see \cite{Davis, Tverskoi67, Kulsrud71,  Achterberg81,Schlickeiser89}  and the reviews by \cite{Melrose2009, Miller97, Petrosian12}).

A more general framework to treat the acceleration and transport of particles from stochastic em fields is the {\bf Fokker-Planck} (FP) transport equation \citep{Gardiner94}. The FP is valid when several approximations for the nature of the em fluctuations are valid.  

	 The main assumption is that the statistical properties of the interaction of the charged particles with the fields are dominated by Gaussian distributions. 
	This is in correspondence with the Central Limit Theorem, which requires that all stochastic systems evolve {\bf asymptotically} towards Gaussian statistics provided that (a) many interactions are involved, (b) the change in state in individual interactions is always small, and (c) subsequent changes of state are statistically independent of each other. Besides these necessary conditions for the applicability of the FP equation, also other simplifying assumption for a better tractability of the FP equation are being made, such as (1) the magnetic fluctuations are homogeneous in space, (2) the electromagnetic fields are quasi-static, (3) the interaction has a finite decorrelation time, etc.\ (see more details in the books \citep{Schhlickeiser02Book, Zank14book}). 
Unfortunately, in astrophysical and laboratory plasmas, most of the above assumptions are not valid, yet the FP equation is used extensively, without a proof of its validity. This is especially true when the plasma particles are accelerated to high energies impulsively (e.g.\ in solar flares, coronal mass ejections, or the Earth's magnetotail). The acceleration volume is finite and  the expected fluctuating electromagnetic fields are strong ($\vert \delta \vec{B} \vert \geq  \vert \vec{B}_0 \vert $). In solar active regions the complex magnetic topologies host many null magnetic points which  are randomly distributed  inside the erupting or flaring volume \citep{Aulanier00, Pontin11}. In these cases the interaction of the particles with the strong em disturbances is transient and has no time to lead to Gaussian statistics or to become homogeneous in space \citep{Isliker19}.  Before analyzing the interaction of the particles with the em fluctuations, it is important to understand the evolution of the em waves. With the use of resistive MHD codes one can show that a spectrum of high amplitude electromagnetic fluctuations evolves rapidly and leads to a fragmented current system, where reconnecting current sheets and large amplitude magnetic fluctuations are present, \citep{Dmitruk04, Vlahos04, Arzner04, Isliker17a}, 
 and where the various statistics clearly are non-Gaussian, following largely the paradigm of the stable Levy distributions \citep{Isliker19,Isliker17a}.
In strongly turbulent plasmas the  magnetic fluctuations are  non-collective modes and  cannot be described with a simple dispersion relation $\omega=\omega(\vec{k}).$  The em environment generated from the evolution of large amplitude em fluctuations  is well documented in the current literature and models much better many impulsive astrophysical and laboratory plasmas  (see \citep{Dmitruk04, Zhdankin13, Isliker17a} and the reviews  by \cite{Cargill12,  Karimabadi2014, Vlahos18}). 

We pose in this article the following question: ``Can we analyze   acceleration and transport  of particles in a finite and strongly turbulent magnetized plasma  with the use of the FP transport equation?".

We analyze the transport   of particles in space and energy in a strongly turbulent environment, where particles gain and loose energy stochastically, following the second order Fermi acceleration scenario  \citep{Fermi49}. In section 2 we introduce the main concepts of  stochastic acceleration and the basic tools available to study  non-Gaussian (anomalous) transport. In section 3 we present our model and in section 4 our results. In the final section we discuss the astrophysical implications of our study and summarize the main points and the limitations of our analysis.

\section{Stochastic acceleration and anomalous transport} 
 The simplest way to model the kinetic properties of an ensemble of particles in the presence of em fields is to assume that the system particles and fields are very close to equilibrium and that the particles perform a Brownian motion (random walk) among the scattering fields, with a well defined mean displacement between encounters (mean free path $\delta r \approx  \lambda_{sc}$). Another way to describe mathematically the random walk inside a system with homogeneously distributed scatterers (magnetic fluctuations or magnetic clouds) is to define the probability density of the step size of the random walk $q(\delta r).$ Based on the Central Limit Theorem (CLT), in all large systems the random steps will asymptotically attain a Gaussian distribution. In this case the mean square displacement, when assuming constant time steps, 
 will be 
\begin{equation} \label{normalDiff}
	<(\Delta r)^2>=D_0 t.
\end{equation}
where the transport coefficient for normal transport is $D_0=v \lambda_{sc},$ and where $v$ is the constant velocity of the particles. It is easy to connect this analysis with the derivation of the FP equation (for details see the tutorials \cite{Vlahos08Tut, Bovet15}). 

Many systems are far from equilibrium and violate many of the  assumptions listed above. Transport in systems far from equilibrium is termed  ``anomalous'' or ``non-Gaussian''. A simple way to characterize  the anomalous transport properties is to estimate the mean square displacement $(\Delta r)^2$. It has been shown that the mean square displacement   usually changes  faster or slower with time than in the case of normal diffusion when fluids or plasmas are strongly turbulent,
\begin{equation} \label{andiff}
	<(\Delta r)^2>=D_r t^{a_r}.
\end{equation}
where $a_r$ can be smaller than one, which is called subdiffusion or $1<a_r$, and which is called superdiffusion. When $a_r=2$ the transport is termed ballistic (for a quick overview see \cite{Shlesinger93, Klafter96} or the  review by \cite{Metzler00, Metzler04}). When particles move with constant velocity and are temporarily trapped inside the scatterers, then their transport is subdiffusive $(a_r<1)$. Superdiffusion means that the particle trajectories involve Levy flights, besides regular steps, during the motion in space \citep{Zimbardo15}, superdiffusive transport regimes are thus based on Levy-type statistics if the velocity is assumed to remain constant, i.e.\ the probability distributions of the step sizes are characterized by power law tails, rather than following Gaussian statistics.  
The theoretical description of  anomalous transport involves the use of a variety of tools and models, like 
the estimate of Hurst exponents, and 
the construction of fractional transport equations
(e.g.\ \cite{Eule12}), not always though it is  clear which model is the most appropriate to describe a specific physical system. 

Anomalous transport has been found in a large variety of physical systems \citep{Zimbardo10, Perrone13}.
In space and astrophysical plasmas, non classical transport and the coupling of particles with strong em fields is common. The wealth of transport studies during the recent decades has shown the existence of a variety of regimes that differ from the classical QL regime \citep{Zimbardo10, Perrone13}.

All the above studies decoupled the spatial transport from the acceleration of particles. They assume that inside the acceleration volume the particles diffuse only in velocity space, and when they escape 
from the acceleration region
they travel in an environment where only spatial transport is analyzed \citep{Perri07}. In the well known stochastic (second order) Fermi acceleration, used extensively in space and astrophysical plasmas, the spatial transport and the acceleration of particles are coupled. Scatterers (magnetic clouds or large amplitude magnetic disturbances) are not just changing the direction of propagation  or the steps of the ``walk'' of the particles, but they also change stochastically their velocity \citep{Fermi49}.  If $W$ is the kinetic energy of a particle and the particle moves with  relativistic velocity $v$ and the scatterers (``magnetic clouds'') move with mean speed $V$, which is much smaller than the speed of light, then
    the energy loss or gain of the particles interacting with the scatterers is
        \begin{equation}\label{energyF}
	        \frac{\delta W}{W}\approx\frac{2}{c^2}(V^2-\vec{V} \cdot \vec{v}) ,
         \end{equation}
    where for overtaking collisions $\vec{V} \cdot \vec{v}>0$ and the particles lose energy,
    for head on collisions $\vec{V} \cdot \vec{v}<0$ and the particles gain energy.
     The rate of energy gain is estimated through the relation
        \begin{equation}\label{rate}
	        \frac{d W}{d t}=\frac{W}{t_{\rm acc}},
        \end{equation}
    where
        \begin{equation}\label{timeacc}
	        t_{\rm acc}= \left (\frac{3c}{4V^2} \right ) \lambda_{\rm sc},
	    \end{equation}
    and $\lambda_{\rm sc}$ is the mean free path the particles travel between the scatterers  and $t_{acc}$ is the acceleration time  \citep[see][]{Longair11}.  On  
    assuming a uniform distribution of the scatterers inside the acceleration volume with density $n_{\rm sc}$, the mean free path is $\lambda_{\rm sc} \approx \frac{1}{ \sqrt[3]{n_{\rm sc}}}.$ The assumption that the scatters are distributed uniformly in space is a {\bf strong} assumption for turbulent systems, since the latter  tend to be highly anisotropic and thus also the distribution of the scatterers may not be isotropic or form a fractal structure. We assume that the particles are not trapped inside the scatterers, i.e.\ {\bf their interaction is instantaneous}. A key parameter for the efficient acceleration of particles inside a turbulent volume is the 
    escape time ($t_{esc}$) from the energy release volume.  
    In second order Fermi acceleration, 
    the spectral index of the high energy particles depends on the ratio $t_{acc}/t_{esc}$
       \begin{equation}
    	f(W) \approx W^{-k}
    \end{equation}
     where  $ k = 1 + t_{\rm acc}/t_{\rm esc}$
 \citep{Longair11}. Therefore the correct estimate of the transport properties inside a finite acceleration volume is very important for the understanding of many transient astrophysical sources. 
  
   The  transport of particles inside an acceleration volume has been analyzed using analytical and numerical tools by several authors \citep{Bouchet04, Tsironis05, Stawicki05, Perri07a, Pisokas16}. \cite{Bouchet04} analyzed the stochastic Fermi acceleration using a simplified model, and they concluded that both the velocity and position mean square displacements 
   \begin{equation}
   	<( \Delta r(t))^2> \approx t^{a_r}
   \end{equation}
   \begin{equation}
   	<(\Delta v(t))^2> \approx t^{a_v}
   \end{equation}
   are superdiffusive, and the probability distributions 
   of the positions and velocities are non-Gaussian. 
   
    The role of anomalous transport upstream of a shock during Diffusive Shock Acceleration has been analyzed in detail \citep{Perri07, Perri08, Perri09, Zimbardo13}.  It was shown that  electron transport upstream of the shocks that are associated with corotating  interaction regions, detected by the Ulysses spacecraft in the solar wind at 4-5 AU, is superdiffusive with $a_r \approx 1.1-1.7.$  The astrophysical implications  of  superdiffusive shock acceleration are discussed in several articles and reviews  \citep{Perri12,Zimbardo15}.

\section{Our model}\label{the_model}

We construct a 3D grid $(N \times N \times N)$ 
with grid-size $\ell=L/(N-1)$, 
and with linear extent $L$. 
Each grid point is set to either \emph{active} or \emph{inactive}, i.e.~it is a scatterer or not. Only a small fraction $R = N_{\rm sc}/N^3$ of the grid points are set to active (5-\SI{15}{\percent}). We can define the density of the scatterers as $n_{\rm sc} = R \times N^3/L^3$,
and the mean free path of the particles between scatterers can be determined as $\lambda_{\rm sc}=\ell/R.$
When a particle (an electron or an ion) encounters an active grid point, it renews its kinetic energy  state 
according to
Eq.\ \ref{energyF}, as it holds for stochastic Fermi acceleration. It then moves in a random direction along the grid
 \begin{equation} \label{spacestep}
\vec{r}^{\;n+1} =\vec{ r}^{\;n} +\delta \vec{r}^{\; n}
\end{equation}
with its renewed velocity 
\begin{equation} \label{Velstep}
	\vec{v}^{\; n+1} =\vec{v}^{\;n} +\delta \vec{v}^{\; n}
\end{equation}
during the  step $n$,
until it meets another active point 
or exits the grid. The minimum distance between two scatterers is the grid size ($\ell$). The time between two consecutive scatterings is $\delta t^{\;n} =\vert \delta \vec{r}^{;n} \vert  /\vert \vec{v}^{\;n} \vert .$  At time $t = 0$ all particles are located at random positions on the  grid. The injected distribution $n(W, t=0)$ is a Maxwellian with temperature $T$. The initial direction of motion of every particle is selected randomly.
 The parameters used in this article are related to the typical plasma parameters in the low solar corona. We choose the strength of the magnetic field to be $B = \SI{100}{G}$, the density of the plasma $n_0 = \SI{e9}{cm^{-3}}$, and the ambient temperature around $\SI{100}{eV}$. The Alfv\'en speed is $V_A \simeq \SI{7e8}{cm/sec}$, so $V_A$ is comparable with the thermal speed of the electrons. The energy increments are of the order of $(\delta W/W) \approx (V_A/c)^2\sim \num{e-4}$ (see Eq.~\ref{energyF}), and the length $L$ of the simulation box  is \SI{e10}{cm}. When not stated otherwise, we consider the grid to be open, i.e.\ particles can escape from the acceleration volume in case they reach any of the boundaries of the grid, at $t = t_{\rm esc}$, which of course is different for each particle that escapes. We usually assume 
 that only $R = \SI{10}{\percent}$ of the $N^3 = 601^3$ grid points are active.

\begin{figure}
\includegraphics[width=8cm]{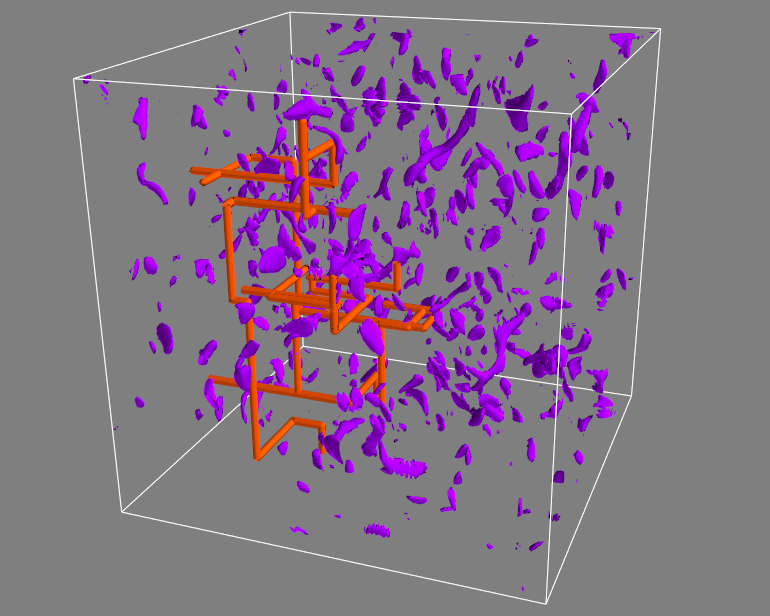}\
\includegraphics[width=8cm]{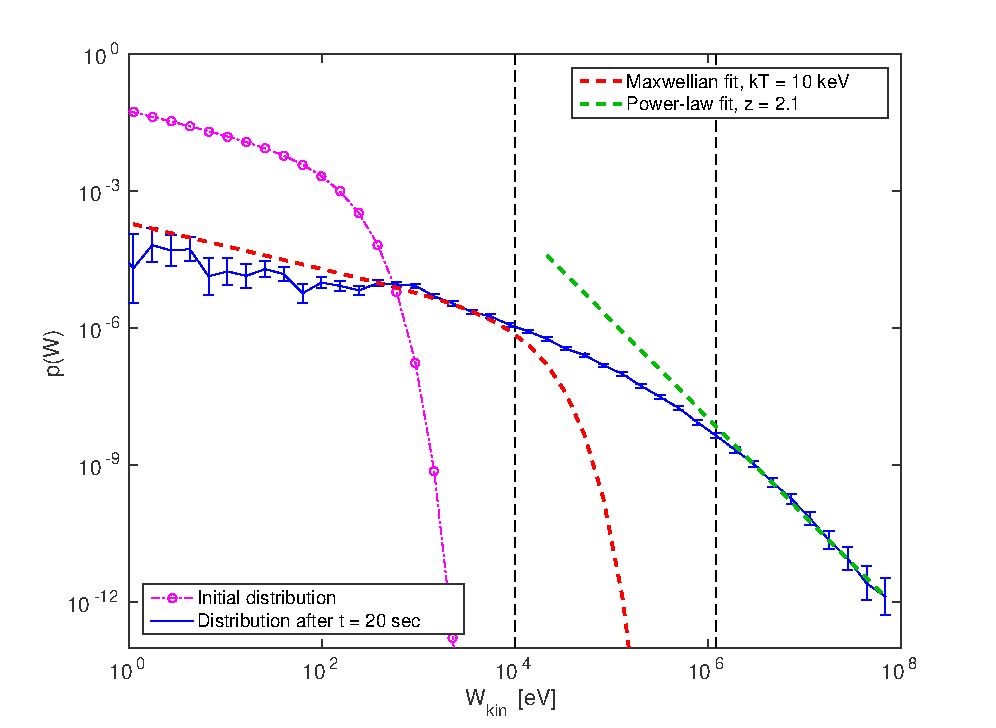}	%
	\caption{(a) Illustration of a trajectory of a typical particle (magneta line) inside the grid. Active points are represented by coherent structures, as commonly seen in large scale MHD simulations. The particle starts at a random grid-point, moves along a straight path on the grid till it meets an active point, and then it moves into a new random direction, and so on, until it eventually exits the simulation box. (b) Energy distribution at $t=0$ sec and $t=20$ sec (stabilized) for the electrons remaining inside the box. Stochastic Fermi Acceleration from the large scale magnetic didturbances  formed insid a strongly turbulent plasma,  as seen in the simulation box above, heats and accelerates the particles from the thermal pool\citep{Pisokas16}. }\label{f:3dBox}
\end{figure}

Large scale magnetic fluctuations  are formed inside unstable  complex magnetic topologies (see Fig.~\ref{f:3dBox}a). In several MHD numerical simulations the formation of coherent structures (current sheets) is also present (see for example Fig.\ 1 in  \cite{Dmitruk04} or Fig.\ 1 in \cite{Isliker17a}). In this article, we exclude, for the sake of simplicity, the presence of unstable current sheets and their contribution to stochastic Fermi acceleration \citep{Vlahos16, Pisokas18}. A typical trajectory of a particle inside the simulation box is displayed in Fig.~\ref{f:3dBox}, the particle moves along the grid on straight lines until it encounters a scatterer, which affects its energy and direction of motion (see Eq.~\ref{energyF}). 
The motion of the particle thus consists of the combined random walk in 
position and velocity space,
with random gains and losses of energy, before exiting the simulation box. 
The mean free path is 
$\lambda_{\rm sc} = \ell/R \simeq \SI{2 e8}{cm}$, which coincides with the value estimated numerically by tracing particles inside the simulation box.

Using the numerical model  presented above, \cite{Pisokas16} showed that the model reproduces all the known analytical results of the stochastic Fermi acceleration. They also showed that the stochastic Fermi energization can reproduce the well known energy distribution of  astrophysical plasmas, where heating of the bulk of the plasma (here with energies $E<10 \; KeV$) and acceleration of the energetic particles (here for energies above $1 \; MeV$; see Fig.\ \ref{f:3dBox}(b)) takes place. The density of the scatterers (which is equivalent to the mean free path $\lambda_{\rm sc}$ of the interaction of the particles with the scatterers) controls 
the evolution of the energetic particles  
and the heating. The energy distribution reaches an asymptotic state on a time scale comparable to the acceleration time $t_{\rm acc}$ \citep[see Fig 3 in][]{Pisokas16}. Similar results have been reported on the interaction of ions with a spectrum of Alfv\'en waves or of electrons with a spectrum of whistler waves \citep[see][]{Miller90}. When the energy distribution reaches the asymptotic state, the mean escape time of the particles $t_{\rm esc} \sim 8$ sec is close to the acceleration time $t_{\rm acc} \sim 9$ sec. The index of the power-law of the particles in the energetic tail 
then agrees very well with the simple formula derived by Fermi, $k=1+t_{\rm acc}/t_{\rm esc} \sim 2$, see Fig.\ \ref{f:3dBox}(b).

\citet{Parker58, Ramaty79} and \citet{Blandford87} analyzed the interaction of electrons and ions with large amplitude magnetic perturbations, which they assumed to be {\bf hard spheres} in order to be able to obtain analytical results. In the hard sphere approximation, the mean energy increase is
\begin{equation} \label{meanhs}
	 <W(t)> \sim t^2.
\end{equation}
The energy distribution is obtained as an analytical solution of the Fokker Planck equation.
For particles with low energy $(W<<mc^2)$ it can be approximated with a Maxwellian distribution, and for the high energy, relativistic particles ($W>>mc^2$), the solution is
\begin{equation} \label{heparticles}
	f(W) \sim W^{1/2- (1/2)(9+12[t_{acc}/t_{esc}])^{1/2}} .
\end{equation}
Thus also in the hard sphere approximation, the escape time $t_{esc}$, which is closely related with  the transport of particles inside the acceleration volume, plays a key role in the estimation of the power law index of the high energy particles. For infinite acceleration regions and $t_{esc} \rightarrow \infty$, the energy distribution function is $f(w) \sim W^{-1}$ and for $t_{acc}/t_{esc} \sim 1$, it takes the form $f(W) \sim W^{-2}.$
Therefore the results reported by Fermi in his original article would have to be modified for non relativistic  or relativistic particles, if the analysis is based on the assumption of hard spheres, and $t_{esc}$ should be estimated correctly from the dynamics inside the acceleration volume. 

\section{Results} 

\subsection{Spatial Diffusion}\label{spatialdiff}

We start our simulation by putting the particle with index $j$ at a random position $\vec{r}_{0j}$  in the 3D simulation box shown in Fig.\ \ref{f:3dBox}a.  
We follow the spatial evolution of the particle for $n_j$ time steps until its individual time $t_j^n= \sum_{i=1}^n \delta t_j^i$ (see Sec.\ \ref{the_model}) reaches $t_{final}$, which is chosen shorter than the mean time the particles need to escape from the acceleration volume \citep{Pisokas16}. The particle positions are monitored at several hundred, predefined and equi-spaced monitoring times $t^m\ (m=1,...M)$ in the interval $[0,t_{final}]$, such that all particles are kept track of at equal times.
	The particle reaches the position $\vec{r}_j^m=\vec{r}_j(t^m)$ at time $t^m$ and its displacement  from its initial position is $\Delta \vec r_j^{\;m}=\vec{r}_j^{\;m} - \vec{r}_{0j}$. 
We collect the  $\vec{r}_j^m$  for $N_p$ particles and we estimate the
mean square displacement as
$$<(\Delta r^m)^2>= \frac{1}{N_p}\sum_{j=1}^{j=N_p}(\Delta r_j^m)^2.$$
The typical number of test particles used are several hundreds of thousands for all the simulations reported in this article.

\subsubsection{Normal transport}
\begin{figure}
	\includegraphics[width= 8cm]{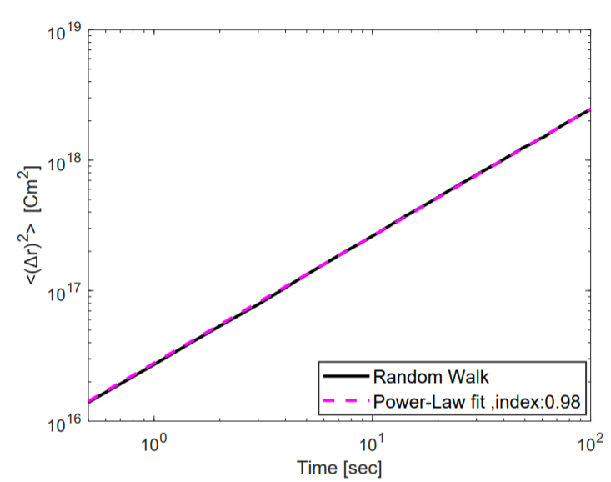}
\includegraphics[width=8cm]{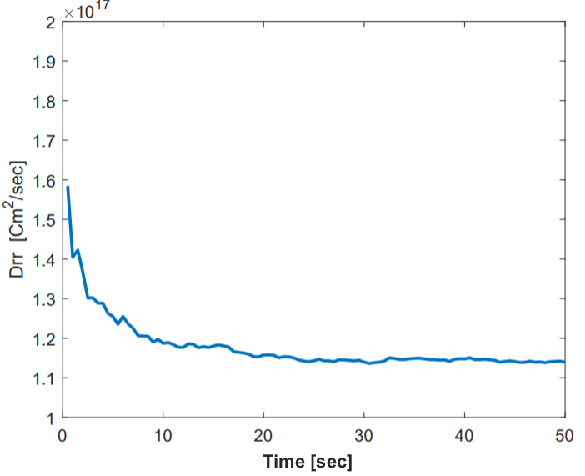}	 
\caption{(a) The mean square displacement of the particles as a function of time, for the classical random walk without energization of the particles. The power law fit has a slope equal to one. (b) The diffusion coefficient $D_{rr}$ as a function of time.} \label{brownian}
 \end{figure}

As a consistency test, we assume here that the particles diffuse inside the simulation box without exchanging energy with the scatterers, so that their motion is a classical random walk and their transport properties have been discussed already (see Eq.\ \ref{normalDiff}).  We plot in Fig.\ \ref{brownian}a the  mean square displacement  as a function of time.  The slope of the power law fit is indeed one ($a_r=1$), as it expected for  normal transport (Eq.\ \ref{normalDiff}).

From the same data, we can also estimate the diffusion coefficient  
\begin{equation} \label{msqdisp} 
    D_{rr}=\frac{1}{2}\frac{<{(\Delta}{r}^{n})^{2}>}{t},
    \end{equation}
see Fig.\ \ref{brownian}b.
The value of $D_{rr}$ can also be estimated from the relation $D_{rr} \sim 3 \lambda_{sc} \cdot v_{the} \sim 10^{17} \rm{cm^2/sec}$  (see Eq.\ 8 in \cite{Vlahos08Tut}) for a classical random walk, which agrees very well with the simulation results in Fig. \ref{brownian}b.

\subsubsection{Anomalous transport}

\begin{figure}
\includegraphics[width=8cm]{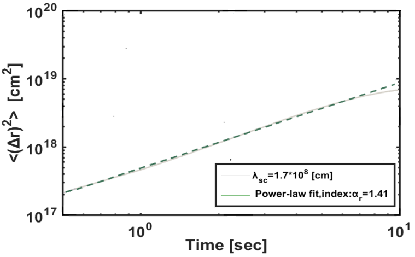}
\includegraphics[width=8cm]{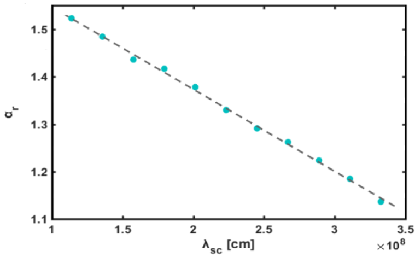}\\
\includegraphics[width=8cm]{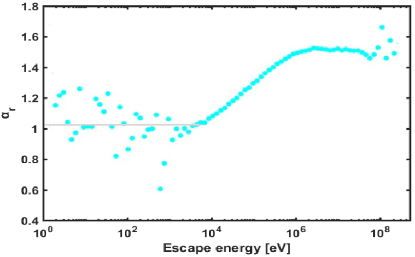} 
	 	\caption{(a) Mean square displacement of the accelerated electrons, the index $a_r$ is 1.6 for  $\lambda_{sc} \sim 2\cdot 10^8 {\rm cm}.$ (b) The index $a_r$ as a function of $\lambda_{sc},$ (c) The index $a_r$ as a function of  the energy
	 	at final time or the energy with which the electrons escape from the acceleration volume, for $\lambda_{sc} \sim 1.7 \times 10^8 \; cm$}\label{antr}
	 \end{figure}

We now turn to the transport properties of the electrons energized inside the simulation volume, 
with the set up presented above, and  assuming that the particles are interacting with active scatterers that realize the stochastic Fermi process. The electrons thus gain or lose energy stochastically, following Eq.\ \ref{energyF}. The  mean square displacement as a function of time for a typical value of $\lambda_{sc}$ is shown in Fig.\ \ref{antr}a, 
it obviously 
follows Eq.\ \ref{andiff}, 
with power-law
index $ a_r \sim 1.41 $ for $\lambda_{sc} \sim 2 \cdot  10^8 cm$,
diffusion thus is clearly anomalous.
 
 In Fig.  \ref{antr}b) we show the index $a_r$ as a function of $\lambda_{sc}$ for typical values of the latter for the solar corona \citep{Kontar17}. We conclude from the relation of $a_r$ with  $\lambda_{sc}$ that electrons diffuse faster inside the acceleration volume and escape from the acceleration volume  in shorter times when the mean free path $\lambda_{sc}$ is smaller. The superdiffusive characteristics of the electrons are related with  the fact that acceleration and spatial transport are coupled \citep{Bouchet04}, which is a rather large departure from the picture of anomalous  transport of particles with constant velocity, discussed earlier in section 2 (see the also the review \cite{Metzler00}). In other words, there are no spatial Levy flights involved in the superdiffusive spatial transport.


The degree of energization of the test particles depends strongly  on their  trapping  inside the turbulent volume.   
To illustrate this, we consider
the relation between the spatial mean  square displacement and its characteristic scaling index $a_r$ with the energy at final time or the energy with which the electrons escape from  the acceleration volume.  We divide the particles according to their final or escape-energy into (logarithmically equi-spaced) bins, and the index  $a_r$ is computed separately for the particles belonging to each bin, using information from their travel history from $t=0$ up to $t_{esc} \sim t_{final} \sim 10 {\rm sec}.$    
Fig.\ \ref{antr}c shows the scaling index $a_r$ as a function of the final or escape energy.  
Electrons with energy $E < 10^4 keV$ are transported normally ($a_r \sim 1$) inside the acceleration volume and can be studied with the Fokker-Planck equation. The particles in the high energy tail   $(E >1 MeV)$, on the other hand, move in a superdiffusive way, 
with  $a_r \sim 1.6$ (see Fig.\ \ref{antr}c).	 The electrons that escape with energies between $10^4 \; keV$ and $1 \; MeV$ are gradually becoming super-diffusive  and $a_r$ increases linearly with the escape energy from 1 to 1.6.

 \begin{figure}
\includegraphics[width=9 cm]{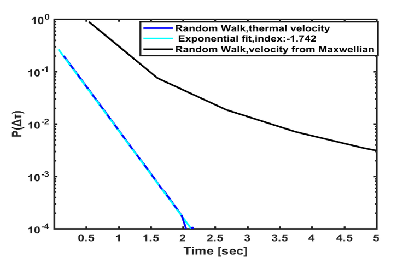}
\includegraphics[width=8 cm]{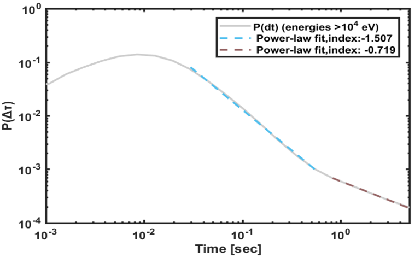}
	\caption{ (a) The probability distribution $P(\Delta t) $ of the waiting times between collisions of the particles with passive scatterers, for the case with random (Maxwellian) velocities and for the case with equal velocity for all particles (equal to the thermal velocity).
	(b) Waiting time distribution $P(\Delta t) $ for the interaction with active scatterers (stochastic Fermi acceleration). }\label{Pwtf}
	 \end{figure}

The waiting time probability distribution $P(\Delta t)$ of the time intervals $\Delta t$ in-between collisions of the particles with the scattering centers is an important parameter of the interaction of the particles with the large scale magnetic disturbances. If the scattering centers are passive and the particles execute a classical random walk with constant speed before exiting the simulation box, we expect the waiting time probability distribution to be of exponential shape,
\begin{equation} \label{Pwt}
	P(\Delta  t)=ae^{-a(\Delta t-t_0)}
\end{equation}
as long as the scatterers are uniformly distributed as in our set up.

	Here $1/a \approx \lambda_{sc}/v_{th}$ is the mean time between collisions with the scatterers, with $v_{th}$ the thermal velocity of the initially injected Maxwellian distribution, and $t_0= \ell/v_{th} $ is the smallest possible waiting time ($\ell$ is the grid-spacing).

	From the parameters used in the numerical simulations, we can estimate $1/a \approx \lambda_{sc}/v_{th}= 2\times 10^8 cm/ 4 \times 10^8 cm/sec \approx  0.5 sec$ and $t_0= \ell/v_{th} \approx 0.04\,$sec. In Fig.\ \ref{Pwtf}a we show $P(\Delta t)$ as a function of the waiting-time for the case of passive scatterers, and the distribution agrees very well with Eq.\ \ref{Pwt} 
	for the case where all particles have the same velocity (equal to the thermal velocity).  An exponential fit yields $a \sim 1.7\, sec^{-1}$, a value close to the estimate based on the model parameters. For the case where the particles have random (Maxwellian) velocity and still interact with passive scatterers (Fig.\ \ref{Pwtf}a), we find a clear deviation from the exponential shape, the distribution decays slower than exponential. Thus, the just velocity-wise in-homogeneous random walk already shows increased dynamic complexity.
	Fig.\ \ref{Pwtf}b shows $P(\Delta t)$ for the case of active scatterers, the distribution now exhibits a double power-law, with index $1.5$ for the intermediate waiting times and index $0.7$ for the large waiting times.
We note that if we would prescribe this kind of waiting time distribution and the particles would move with constant velocity, then the resulting transport would be of sub-diffusive nature.
In this sense, we must consider that the power-law shaped waiting time distribution  is an indication of the complexity that is caused by the synergy of spatial and active energy transport inside the stochastic Fermi accelerator.  
Our result also clearly departs from the waiting time distribution  $P(\Delta t$) in the work of \cite{Bouchet04}, who present an analytical model of stochastic Fermi acceleration and who also find that transport is clearly anomalous, even though they actually prescribe a waiting time distribution with an exponential tail for their results to hold.

\begin{figure}
	\includegraphics[width=8 cm]{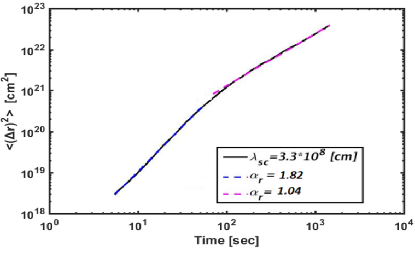}
	 \caption {The mean square displacement as a function of time, for the case of periodic boundary conditions, with two power law fits, one for $t<100\,$sec with $a_r \approx 1.8$, and one for $t>100\,$sec with $a_r \approx 1.0$. }\label{wperiod}
	 \end{figure}

Keeping the setup used above with open boundary conditions,  we now impose periodic boundary conditions, i.e.\ when an electron reaches any of the boundaries of the simulation box it re-enters the box from the opposite boundary.  In this case, we can distinguish two parts in the evolution of the mean square displacement: (a) A transient phase, lasting 100 sec (for plasma parameters related to the low solar corona, as before), during which $a_r \sim 1.8$ for  $\lambda_{sc} \sim 3 \cdot 10 ^8 cm$   (see Fig.\ \ref{wperiod}). (b) A longer and asymptotic phase, where the mean square displacement shows the scaling of normal diffusion. 

It follows that in the periodic case, stochastic Fermi acceleration asymptotically attains Gaussian statistics for the probability distribution of the position of the particles, and the Fokker Planck transport equation can be expected to describe well the spatial evolution of the particle distribution. 
The asymptotic power law index $a_r=1$ of the mean square displacement is in agreement with the prediction of Eq.\  \ref{heparticles} when $t_{esc}$ practically is infinity, as it holds for periodic systems \citep{Parker58,Ramaty79}.

\begin{figure}
	\includegraphics[width=8 cm]{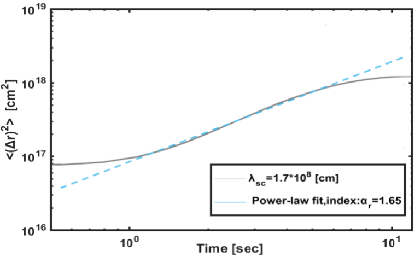}
	\includegraphics[width=8 cm]{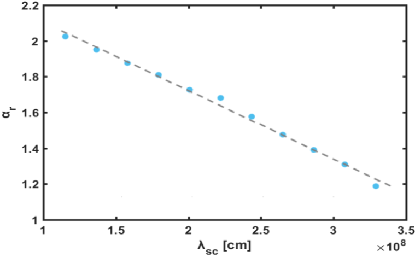}
	\caption{ (a) The  mean square displacement as a function of time for ions, together with a power law fit.
	(b) The index $a_r$ as a function of  $\lambda_{sc}$.}\label{aRvstimeIons}
	 \end{figure}

Our analysis so far was concentrated on the heating and acceleration of  electrons. The transport properties of ions are very similar with the ones reported here for electrons when $\lambda_{sc}$  is the same. In the case of open boundary conditions, there are no significant differences from the behavior  of the electrons, apart from the fact that the required time needed for the energy distribution  to reach the asymptotic state is much longer. The spatial diffusion process is superdiffusive (see Fig.\ \ref{aRvstimeIons}a), and the index $a_r$  decreases when  the mean free path of the ions ${\lambda_{sc}}$ become larger and approaches 1 when $\lambda_{sc}$ approaches  $L,$ the size of the acceleration volume,  
the ions thus also  diffuse and escape faster as $\lambda_{sc}$ becomes smaller (see Fig. \ref{aRvstimeIons}b). The characteristics of $a_r$ as a function of the final or escape energy of the ions 
are also similar with the ones reported for the electrons.
	 
\subsection{Energy transport}

The energy  transport is closely related with the spatial transport when the particles are inside a  finite size stochastic Fermi accelerator. The interaction of the particle $j$  with an active scatterer renews its energy 
$$W_j^{n+1}=W_j^n+\delta  W_j^n$$
where $\delta W_j^n$ is given by Eq.\ \ref{energyF}, and $n$ counts the number of energization events. 
For the mean square displacement, we use the same set of predefined monitoring times $t^m$ as described in Sec.\ \ref{spatialdiff} and keep track of the energies $W^m_j$ at these times, $W^m_j = W_j(t^m)$, so that the mean is evaluated at equal times for all particles.   We then define $\Delta  W_j^m=(W_j^m - W_j^0)$, the energy displacement  from the energy with which the particle was initially injected ($W_j^0$)  into the acceleration volume.

	The mean  displacement of the energy is defined through the relation 
	\begin{equation}\label{enertr}
	<\Delta  W>(t^m)\equiv <\Delta  W^m>=\frac{1}{N_p}\sum_{j=1}^{j=N_p} \Delta W_j^m.
	\end{equation}
	and  the mean square displacement is given as
	\begin{equation}\label{sqenergytr}
	<(\Delta  W)^2>(t^m) \equiv <(\Delta  W^m)^2>=\frac{1}{N_p}\sum_{j=1}^{j=N_p} (\Delta W_j^m)^2.
	\end{equation}

The coefficients of the Fokker-Planck equation in energy-space would be 
\begin{equation} \label{contr}
	 F= \frac{<\Delta W^m>}{\Delta t}\end{equation}
for the  convective term (systematic acceleration) and  
\begin{equation} \label{Difftr}
	 D=\frac{<(\Delta W^m)^2>}{2\Delta t}\end{equation}
for the diffusive term. 
	
  \begin{figure} 
	\includegraphics[width=8 cm]{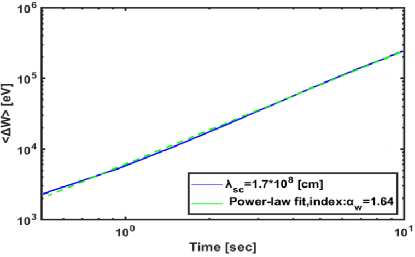}
	\includegraphics[width=8 cm]{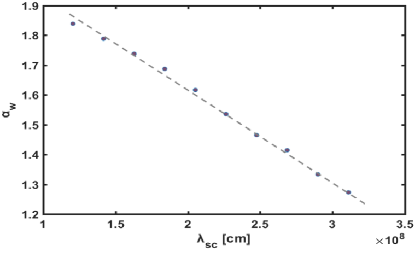} \\
	\includegraphics[width=8 cm]{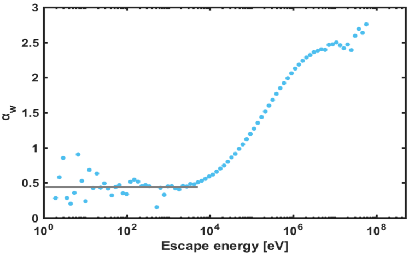}
	\caption{ (a) The mean displacement in energy of the electrons as a function of time, together with a power-law fit. The slope of the power law fit is $a_W \sim 1.64$ for $\lambda_{sc} \sim 2 \cdot 10^8 cm.$ (b) The index $a_W$ as a function of $\lambda_{sc}.$ 
	(c) The scaling index $a_W$ as a function of the final energy or the energy with which the electrons  escape from the acceleration volume.}\label{aRenergy}
	 \end{figure}

	Assuming that the mean  displacement  of the energy in general  can be expressed as $ <\Delta W>(t) = D_{W} t^{a_W},$ we estimate $a_W$ from the slope of the mean energy displacement as a function of time in logarithmic representation. In Fig.\ \ref{aRenergy}a   we show $<\Delta W>$ as a function of time,
the mean displacement indeed shows a power-law scaling,
and a power law fit yields $a_W \sim 1.64$  for $\lambda_{sc} \sim  2 \cdot 10^8 cm$.
Fig.\ \ref{aRenergy}b presents $a_W$ as a function of $\lambda_{sc}$, 
the mean displacement of the particles in energy becomes larger as $\lambda_{sc}$ becomes smaller, i.e.\ the convective (systematic) type of motion gets enhanced. 
In Fig.\ \ref{aRenergy}c we explore $a_W$ as a function of the
final energy or the energy with which the particles escape from the acceleration volume. 
For the electrons with energy smaller the $10 keV$ the convective transport 
is much reduced, in contrast to the high energy particles, where the scaling index reaches an asymptotic value close to $a_W\sim 2.5$. The convective transport for energies between $10 keV$ and $1 MeV$ is changing linearly with the energy from $a_W \sim 0.4$ to $a_W \sim 2.5$.

	  \begin{figure}
\includegraphics[width=8 cm]{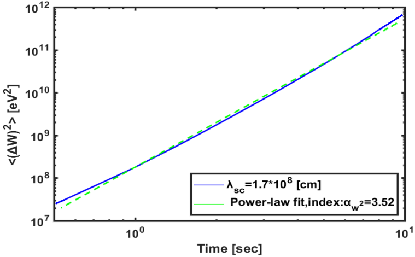}
\includegraphics[width=8 cm]{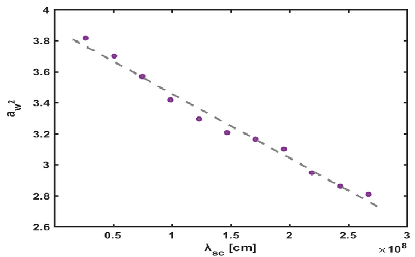}	 \\
\includegraphics[width=8 cm]{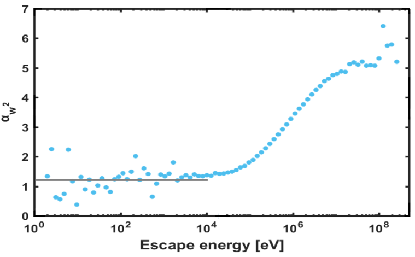}
	\caption{ (a) The mean square displacement in energy for electrons as a function of time. The index of the power law fit is $a_{W^2} \sim 3.52$ for $\lambda_{sc} \sim 2 \cdot 10^8 cm$. (b) The index $a_{W^2}$ as a function of $\lambda_{sc}.$  (c) The scaling index $a_{W^2}$ as a function of the final energy or the energy with which the electrons  escape from the acceleration volume.}\label{aRenergysq}
	 \end{figure}

	We now turn to the mean square displacement in energy, which can generally be expected to have the functional form $ <(\Delta W)^2>(t) = D_{W^2} t^{a_{W^2}}$. 
	Fig.\ \ref{aRenergysq}a  shows
	$<(\Delta W)^2>$ as a function of time,
and by a power law fit we find $a_{W^2} \sim 3.52$  for $\lambda_{sc} \sim  2 \cdot 10^8 cm.$ Transport in energy inside the stochastic Fermi accelerator is also superdiffusive.  In Fig.\  \ref{aRenergysq}b we show $a_{W^2}$ as a function of $\lambda_{sc}$,  diffusion of particles in energy  becomes faster as $\lambda_{sc}$ becomes smaller and the acceleration time also gets shorter (see Eq.\ \ref{timeacc}).  In Fig.\ \ref{aRenergysq}c we show  $a_{W^2}$ as a function of the 
final energy or the energy with which the particles escape from the acceleration volume. 
For the electrons with energy smaller the $10 keV$ the transport 
is normal, and for the high energy particles it is super-diffusive,  the scaling index reaches an asymptotic value close to $a_{W^2}\sim 5$. The transport for energies between $10 keV$ and $1 MeV$ changes linearly with the energy from normal to superdiffusive.

	In order to compare the importance of convection ($F$) and diffusion ($D$) in a FP equation,  
	one actually has to compare $\tau F$ and $\sqrt{2\tau D}$, with $\tau$ a small time-step, of the order of the mean waiting time. 
	From Fig.\ \ref{aRenergy} and Fig.\ \ref{aRenergysq} it then follows that convection and diffusion are of rather similar importance for the dynamics in energy space. Convective effects must be attributed to the fact that Fermi acceleration is slightly biased towards increasing the energies rather than decreasing them.
Given that the transport property of the low energy (heated) particles is normal, and it is anomalous for the high energy particles, another important result from this section is that the Fokker Planck equation is the proper transport equation for the evolution of the low energy particles, while it clearly is an inappropriate tool for modeling the dynamic evolution of the high energy particles.

\section{Discussion and Summary}

For several decades, stochastic Fermi acceleration (or second order acceleration or stochastic turbulent acceleration) was analyzed, almost exclusively, with the use of the QL transport equation in energy \citep{Achterberg81, Melrose2009, Miller90,  Petrosian12}. The time particles remain inside the acceleration volume, $t_{esc}$, was never calculated precisely, and it was used as a free parameter. In many astrophysical and space applications, stochastic Fermi acceleration was excluded as a potential acceleration mechanism, based on the analysis of the QL transport equation as as ``weak and second order process" compared to shock acceleration, which was named a ''first order process". We have already outlined in the introduction the weaknesses of the QL approach and the deficiency of the weak turbulence theory to accelerate electrons and ions efficiently. We believe that the ideas put forward  by Fermi in his article in 1949 have nothing to do with the results derived from the QL transport equation, since Fermi was referring to large scale magnetic disturbances (``magnetic clouds'')  and not to weak turbulence and wave-particle interactions \citep{Kulsrud71}. 

In a recent article by \cite{Pisokas16}, we pose the question: What happens to stochastic Fermi acceleration in a strongly turbulent environment, where the magnetic fluctuations, serving as scatterers, are very large and their interaction with particles follows the formalism proposed initially by Fermi \citep{Fermi49}.

In this article, we have shown that the spatial and energy transport of the high energy particles inside the stochastic Fermi accelerator are both superdiffusive, and 
the time the high energy particles need to escape from the acceleration volume is correlated with the mean free path of the particles between the scatterers. On the other hand, the relation between the transport properties of particles in space and energy plays a very crucial role for the power law index of the high energy tail and generally for the energy distribution inside a finite and impulsive accelerator.

Our main results in this article are:
\begin{enumerate}
	\item The spatial transport of high energy electrons (higher than 1 MeV) inside a finite size  stochastic Fermi accelerator  is superdiffusive with $a_r \sim 1.6,$ while  the transport of heated low energy (smaller than 10 keV) particles   is normal.
	
	\item The transport of particle depends on the mean free path $\lambda_{sc}.$ As $\lambda_{sc}$	 increases, the interaction of the particles with the large amplitude magnetic fluctuations becomes weaker and the transport properties tend asymptotically to normal diffusion. The escape time $t_{esc}$ is linearly  related with $\lambda_{sc},$	i.e.\ the smaller $\lambda_{sc}$ the faster the particles will escape from the acceleration volume.
	\item The probability distribution of the waiting times between subsequent collisions has a double power-law tail, with index $1.5$ for the intermediate waiting times and index $0.7$ for the large waiting times. 
	\item In a periodic simulation box, the spatial transport properties of the particles tend to normal transport as time increases (for the solar corona the characteristic time is higher than 100 seconds for $\lambda_{sc} \sim 3 \cdot 10^8 cm.$).

	\item 	The scaling index $a_W$ of the mean displacement in energy is linearly dependent on the mean free path $\lambda_{sc}.$ The mean displacement in energy becomes larger  as $\lambda_{sc}$ decreases, i.e.\ convective or systematic increase in energy gets enhanced.

	\item The ions have very similar transport characteristics with the electrons, just on a lower time-scale.
	
	\item The mean square displacement in energy is superdiffusive for the high energy particles and normal for the heated low energy particles. The scaling index $a_{W^2}$ depends linearly on the mean square displacement $\lambda_{sc}$, the smaller $\lambda_{sc}$ the faster is transport in energy.   
	This result also agrees very well with Eq.\ \ref{timeacc}.
	
	\item The use of the Fokker-Planck equation for the study of the spatial and energy transport of the high energy particles is inappropriate, but it is valid for modeling the heating of the low energy particles or for long lasting interactions in periodic systems (with infinite escape time).
	 
\end{enumerate}

The astrophysical implications from our analysis are several.

\begin{itemize}
	\item Stochastic Fermi acceleration has been labeled in most space and astrophysical studies as a ``second order" and ``weak'' acceleration process. Returning to the initial ideas proposed by Fermi for the stochastic acceleration, and given the results reported in this article on the superdiffusion of high energy electrons and ions in energy space, we conclude that stochastic Fermi acceleration can be highly efficient and
	the acceleration time is related with the compactness of the scatterers.
	\item The role of stochastic Fermi processes in the heating of plasmas has been ignored.
	\item The second important criticism on stochastic Fermi accelerators is that the index of the high energy particles is not constant and not always close to 2, as most observations suggest, since it depends on the escape time $t_{esc}$, and, in the weak turbulence theory, on the characteristics of the spectrum of the plasma waves accelerating the particles. This result is also misleading, since the transport of the high energy particles analyzed in this articles is superdiffusive and $t_{esc}$ is closely related with $\lambda_{sc}$. Therefore the acceleration time and the escape time are correlated through $\lambda_{sc}.$ \cite{Pisokas16}, using reasonable  values for the low solar corona for $\lambda_{sc}$, show that the power law index of the high energy particles remains close to 2.  
	\item The Fokker Planck transport equation for the evolution of the energy distributions was extensively used in the space and astrophysical plasma literature for the analysis of the interaction of particles with plasma waves. We conclude in this article  that the transport is non-Gaussian for the interaction of particles with  large amplitude magnetic disturbances, and both the spatial and energy transport must be re-examined using a transport framework different from the Fokker-Planck approach \citep[see e.g.][]{Metzler04}. 
	
		In \cite{Isliker2017b}, we have analyzed \textit{systematic} (first order) Fermi acceleration, in the same frame of a lattice model approach as used here. We there have explicitly shown that the Fokker-Planck approach fails, and only a fractional transport equation is adequate and successful in modeling the transport in energy space (spatial transport was not analyzed in \cite{Isliker2017b}). 
		Given the anomalous transport properties of stochastic Fermi acceleration analyzed here (see e.g.\ Fig.\ \ref{aRenergysq}), we would expect that a fractional transport equation similar to that in  
		\cite{Isliker2017b} and \cite{Isliker17a} would be appropriate also for the case of second order Fermi acceleration in what the energy transport is concerned. Still missing is a combined modeling framework for the doubly anomalous transport in position- and energy-space.

	\item \cite{Perri12} discuss the transport of particles upstream of a shock assuming that the turbulence upstream does  not affect the energetics of the particles. In this articles, we show that  superdiffusive characteristics of transport can be the result of stochastic Fermi acceleration in strong turbulence upstream \citep{Zank15, Garrel18}.

	\end{itemize}

	In strongly turbulent plasmas, large scale magnetic disturbances during explosive or impulsive events are associated with 
	coherent structures (unstable current sheets or shocks), which accelerate particles even more efficiently, being first order processes (see \cite{Isliker17a}). In this article, we simplified the analysis by isolating the stochastic Fermi acceleration from the effects of the coherent structures, avoiding in this way the complexity of the synergy of two processes \citep{Pisokas18,Comisso18, Comisso19}. Strongly turbulent plasmas combine all the well known accelerators studied separately  in the past (second order Fermi, unstable current sheets and turbulent shocks; see \cite{Karimabadi03b,Karimabadi2014}. A systematic study of the transport properties of high energy particles in these mixed environments has only recently been started (see \cite{Vlahos04, Zank15, Matsumoto15, Vlahos16,  Pisokas18, Garrel18, Comisso18,  Comisso19}).


\bsp	

\begin{thebibliography}{}
\makeatletter
\relax
\def\mn@urlcharsother{\let\do\@makeother \do\$\do\&\do\#\do\^\do\_\do\%\do\~}
\def\mn@doi{\begingroup\mn@urlcharsother \@ifnextchar [ {\mn@doi@}
  {\mn@doi@[]}}
\def\mn@doi@[#1]#2{\def\@tempa{#1}\ifx\@tempa\@empty \href
  {http://dx.doi.org/#2} {doi:#2}\else \href {http://dx.doi.org/#2} {#1}\fi
  \endgroup}
\def\mn@eprint#1#2{\mn@eprint@#1:#2::\@nil}
\def\mn@eprint@arXiv#1{\href {http://arxiv.org/abs/#1} {{\tt arXiv:#1}}}
\def\mn@eprint@dblp#1{\href {http://dblp.uni-trier.de/rec/bibtex/#1.xml}
  {dblp:#1}}
\def\mn@eprint@#1:#2:#3:#4\@nil{\def\@tempa {#1}\def\@tempb {#2}\def\@tempc
  {#3}\ifx \@tempc \@empty \let \@tempc \@tempb \let \@tempb \@tempa \fi \ifx
  \@tempb \@empty \def\@tempb {arXiv}\fi \@ifundefined
  {mn@eprint@\@tempb}{\@tempb:\@tempc}{\expandafter \expandafter \csname
  mn@eprint@\@tempb\endcsname \expandafter{\@tempc}}}

\bibitem[\protect\citeauthoryear{{Achterberg}}{{Achterberg}}{1981}]{Achterberg81}
{Achterberg} A.,  1981, \aap, \href
  {http://adsabs.harvard.edu/abs/1981A%26A....97..259A} {97, 259}

\bibitem[\protect\citeauthoryear{Arzner \& Vlahos}{Arzner \&
  Vlahos}{2004}]{Arzner04}
Arzner K.,  Vlahos L.,  2004, Astrophys. J. Lett., 605, L69

\bibitem[\protect\citeauthoryear{{Aulanier}, {DeLuca}, {Antiochos}, {McMullen}
  \& {Golub}}{{Aulanier} et~al.}{2000}]{Aulanier00}
{Aulanier} G.,  {DeLuca} E.~E.,  {Antiochos} S.~K.,  {McMullen} R.~A.,
  {Golub} L.,  2000, \mn@doi [\apj] {10.1086/309376}, \href
  {https://ui.adsabs.harvard.edu/abs/2000ApJ...540.1126A} {540, 1126}

\bibitem[\protect\citeauthoryear{{Blandford} \& {Eichler}}{{Blandford} \&
  {Eichler}}{1987}]{Blandford87}
{Blandford} R.,  {Eichler} D.,  1987, \mn@doi [\physrep]
  {10.1016/0370-1573(87)90134-7}, \href
  {https://ui.adsabs.harvard.edu/abs/1987PhR...154....1B} {154, 1}

\bibitem[\protect\citeauthoryear{{Bouchet}, {Cecconi}  \& {Vulpiani}}{{Bouchet}
  et~al.}{2004}]{Bouchet04}
{Bouchet} F.,  {Cecconi} F.,   {Vulpiani} A.,  2004, \mn@doi [\prl]
  {10.1103/PhysRevLett.92.040601}, \href
  {https://ui.adsabs.harvard.edu/abs/2004PhRvL..92d0601B} {92, 040601}

\bibitem[\protect\citeauthoryear{{Bovet}}{{Bovet}}{2015}]{Bovet15}
{Bovet} A.,  2015, arXiv e-prints, \href
  {https://ui.adsabs.harvard.edu/abs/2015arXiv150801879B} {p. arXiv:1508.01879}

\bibitem[\protect\citeauthoryear{Cargill, Vlahos, Baumann, Drake  \&
  Nordlund}{Cargill et~al.}{2012}]{Cargill12}
Cargill P.,  Vlahos L.,  Baumann G.,  Drake J.,   Nordlund {\AA}.,  2012, Space
  science reviews, 173, 223

\bibitem[\protect\citeauthoryear{{Comisso} \& {Sironi}}{{Comisso} \&
  {Sironi}}{2018}]{Comisso18}
{Comisso} L.,  {Sironi} L.,  2018, \mn@doi [\prl]
  {10.1103/PhysRevLett.121.255101}, \href
  {https://ui.adsabs.harvard.edu/abs/2018PhRvL.121y5101C} {121, 255101}

\bibitem[\protect\citeauthoryear{{Comisso} \& {Sironi}}{{Comisso} \&
  {Sironi}}{2019}]{Comisso19}
{Comisso} L.,  {Sironi} L.,  2019, arXiv e-prints, \href
  {https://ui.adsabs.harvard.edu/abs/2019arXiv190901420C} {p. arXiv:1909.01420}

\bibitem[\protect\citeauthoryear{{Davis}}{{Davis}}{1956}]{Davis}
{Davis} L.,  1956, \mn@doi [Physical Review] {10.1103/PhysRev.101.351}, \href
  {http://adsabs.harvard.edu/abs/1956PhRv..101..351D} {101, 351}

\bibitem[\protect\citeauthoryear{{Dmitruk}, {Matthaeus}  \& {Seenu}}{{Dmitruk}
  et~al.}{2004}]{Dmitruk04}
{Dmitruk} P.,  {Matthaeus} W.~H.,   {Seenu} N.,  2004, \mn@doi [Astrophys. J.]
  {10.1086/425301}, \href {http://adsabs.harvard.edu/abs/2004ApJ...617..667D}
  {617, 667}

\bibitem[\protect\citeauthoryear{{Eule}, {Zaburdaev}, {Friedrich}  \&
  {Geisel}}{{Eule} et~al.}{2012}]{Eule12}
{Eule} S.,  {Zaburdaev} V.,  {Friedrich} R.,   {Geisel} T.,  2012, \mn@doi
  [\pre] {10.1103/PhysRevE.86.041134}, \href
  {https://ui.adsabs.harvard.edu/abs/2012PhRvE..86d1134E} {86, 041134}

\bibitem[\protect\citeauthoryear{{Fermi}}{{Fermi}}{1949}]{Fermi49}
{Fermi} E.,  1949, \mn@doi [Physical Review] {10.1103/PhysRev.75.1169}, \href
  {http://adsabs.harvard.edu/abs/1949PhRv...75.1169F} {75, 1169}

\bibitem[\protect\citeauthoryear{{Gardiner}}{{Gardiner}}{1994}]{Gardiner94}
{Gardiner} C.~W.,  1994, {Handbook of stochastic methods for physics, chemistry
  and the natural sciences}

\bibitem[\protect\citeauthoryear{{Garrel}, {Vlahos}, {Isliker}  \&
  {Pisokas}}{{Garrel} et~al.}{2018}]{Garrel18}
{Garrel} C.,  {Vlahos} L.,  {Isliker} H.,   {Pisokas} T.,  2018, \mn@doi
  [\mnras] {10.1093/mnras/sty1260}, \href
  {https://ui.adsabs.harvard.edu/#abs/2018MNRAS.478.2976G} {478, 2976}

\bibitem[\protect\citeauthoryear{{Greco}, {Perri}  \& {Zimbardo}}{{Greco}
  et~al.}{2010}]{Greco10}
{Greco} A.,  {Perri} S.,   {Zimbardo} G.,  2010, \mn@doi [Journal of
  Geophysical Research (Space Physics)] {10.1029/2009JA014690}, \href
  {http://adsabs.harvard.edu/abs/2010JGRA..115.2203G} {115, A02203}

\bibitem[\protect\citeauthoryear{{Isliker}, {Vlahos}  \&
  {Constantinescu}}{{Isliker} et~al.}{2017a}]{Isliker17a}
{Isliker} H.,  {Vlahos} L.,   {Constantinescu} D.,  2017a, \mn@doi [Physical
  Review Letters] {10.1103/PhysRevLett.119.045101}, \href
  {http://adsabs.harvard.edu/abs/2017PhRvL.119d5101I} {119, 045101}

\bibitem[\protect\citeauthoryear{Isliker, Pisokas, Vlahos  \&
  Anastasiadis}{Isliker et~al.}{2017b}]{Isliker2017b}
Isliker H.,  Pisokas T.,  Vlahos L.,   Anastasiadis A.,  2017b, \mn@doi [The
  Astrophysical Journal] {10.3847/1538-4357/aa8ee8}, 849, 35

\bibitem[\protect\citeauthoryear{Isliker, Archontis  \& Vlahos}{Isliker
  et~al.}{2019}]{Isliker19}
Isliker H.,  Archontis V.,   Vlahos L.,  2019, \mn@doi [The Astrophysical
  Journal] {10.3847/1538-4357/ab30c6}, 882, 57

\bibitem[\protect\citeauthoryear{{Karimabadi} et~al.,}{{Karimabadi}
  et~al.}{2013}]{Karimabadi03b}
{Karimabadi} H.,  et~al., 2013, \mn@doi [Physics of Plasmas]
  {10.1063/1.4773205}, \href
  {http://adsabs.harvard.edu/abs/2013PhPl...20a2303K} {20, 012303}

\bibitem[\protect\citeauthoryear{{Karimabadi} et~al.,}{{Karimabadi}
  et~al.}{2014}]{Karimabadi2014}
{Karimabadi} H.,  et~al., 2014, \mn@doi [Physics of Plasmas]
  {10.1063/1.4882875}, \href
  {http://adsabs.harvard.edu/abs/2014PhPl...21f2308K} {21, 062308}

\bibitem[\protect\citeauthoryear{{Kennel} \& {Engelmann}}{{Kennel} \&
  {Engelmann}}{1966}]{Kennel66}
{Kennel} C.~F.,  {Engelmann} F.,  1966, \mn@doi [Physics of Fluids]
  {10.1063/1.1761629}, \href
  {https://ui.adsabs.harvard.edu/abs/1966PhFl....9.2377K} {9, 2377}

\bibitem[\protect\citeauthoryear{{Klafter}, {Shlesinger}  \&
  {Zumofen}}{{Klafter} et~al.}{1996}]{Klafter96}
{Klafter} J.,  {Shlesinger} M.~F.,   {Zumofen} G.,  1996, \mn@doi [Physics
  Today] {10.1063/1.881487}, \href
  {https://ui.adsabs.harvard.edu/abs/1996PhT....49b..33K} {49, 33}

\bibitem[\protect\citeauthoryear{{Kontar}, {Perez}, {Harra}, {Kuznetsov},
  {Emslie}, {Jeffrey}, {Bian}  \& {Dennis}}{{Kontar} et~al.}{2017}]{Kontar17}
{Kontar} E.~P.,  {Perez} J.~E.,  {Harra} L.~K.,  {Kuznetsov} A.~A.,  {Emslie}
  A.~G.,  {Jeffrey} N.~L.~S.,  {Bian} N.~H.,   {Dennis} B.~R.,  2017, \mn@doi
  [\prl] {10.1103/PhysRevLett.118.155101}, \href
  {https://ui.adsabs.harvard.edu/abs/2017PhRvL.118o5101K} {118, 155101}

\bibitem[\protect\citeauthoryear{Kulsrud \& Ferrari}{Kulsrud \&
  Ferrari}{1971}]{Kulsrud71}
Kulsrud R.~M.,  Ferrari A.,  1971, Astrophysics and Space Science, 12, 302

\bibitem[\protect\citeauthoryear{{Kuramitsu} \& {Hada}}{{Kuramitsu} \&
  {Hada}}{2000}]{Kuramitsu00}
{Kuramitsu} Y.,  {Hada} T.,  2000, \mn@doi [\grl] {10.1029/1999GL010726}, \href
  {https://ui.adsabs.harvard.edu/abs/2000GeoRL..27..629K} {27, 629}

\bibitem[\protect\citeauthoryear{{Longair}}{{Longair}}{2011}]{Longair11}
{Longair} M.~S.,  2011, {High Energy Astrophysics}.
Cambridge University Press

\bibitem[\protect\citeauthoryear{{Matsumoto}, {Amano}, {Kato}  \&
  {Hoshino}}{{Matsumoto} et~al.}{2015}]{Matsumoto15}
{Matsumoto} Y.,  {Amano} T.,  {Kato} T.~N.,   {Hoshino} M.,  2015, \mn@doi
  [Science] {10.1126/science.1260168}, \href
  {http://adsabs.harvard.edu/abs/2015Sci...347..974M} {347, 974}

\bibitem[\protect\citeauthoryear{{Melrose}}{{Melrose}}{2009}]{Melrose2009}
{Melrose} D.~B.,  2009, preprint, \href
  {http://adsabs.harvard.edu/abs/2009arXiv0902.1803M} {} (\mn@eprint {arXiv}
  {0902.1803})

\bibitem[\protect\citeauthoryear{{Metzler} \& {Klafter}}{{Metzler} \&
  {Klafter}}{2000}]{Metzler00}
{Metzler} R.,  {Klafter} J.,  2000, \mn@doi [\physrep]
  {10.1016/S0370-1573(00)00070-3}, \href
  {https://ui.adsabs.harvard.edu/abs/2000PhR...339....1M} {339, 1}

\bibitem[\protect\citeauthoryear{{Metzler} \& {Klafter}}{{Metzler} \&
  {Klafter}}{2004}]{Metzler04}
{Metzler} R.,  {Klafter} J.,  2004, \mn@doi [Journal of Physics A Mathematical
  General] {10.1088/0305-4470/37/31/R01}, \href
  {https://ui.adsabs.harvard.edu/abs/2004JPhA...37R.161M} {37, R161}

\bibitem[\protect\citeauthoryear{{Miller}, {Guessoum}  \& {Ramaty}}{{Miller}
  et~al.}{1990}]{Miller90}
{Miller} J.~A.,  {Guessoum} N.,   {Ramaty} R.,  1990, \mn@doi [\apj]
  {10.1086/169233}, \href {http://adsabs.harvard.edu/abs/1990ApJ...361..701M}
  {361, 701}

\bibitem[\protect\citeauthoryear{{Miller} et~al.,}{{Miller}
  et~al.}{1997}]{Miller97}
{Miller} J.~A.,  et~al., 1997, \mn@doi [J. Geoph. Research]
  {10.1029/97JA00976}, \href
  {http://adsabs.harvard.edu/abs/1997JGR...10214631M} {102, 14631}

\bibitem[\protect\citeauthoryear{{Parker} \& {Tidman}}{{Parker} \&
  {Tidman}}{1958}]{Parker58}
{Parker} E.~N.,  {Tidman} D.~A.,  1958, \mn@doi [Physical Review]
  {10.1103/PhysRev.111.1206}, \href
  {http://adsabs.harvard.edu/abs/1958PhRv..111.1206P} {111, 1206}

\bibitem[\protect\citeauthoryear{{Perri} \& {Zimbardo}}{{Perri} \&
  {Zimbardo}}{2007}]{Perri07}
{Perri} S.,  {Zimbardo} G.,  2007, \mn@doi [\apjl] {10.1086/525523}, \href
  {https://ui.adsabs.harvard.edu/abs/2007ApJ...671L.177P} {671, L177}

\bibitem[\protect\citeauthoryear{{Perri} \& {Zimbardo}}{{Perri} \&
  {Zimbardo}}{2008}]{Perri08}
{Perri} S.,  {Zimbardo} G.,  2008, \mn@doi [Journal of Geophysical Research
  (Space Physics)] {10.1029/2007JA012695}, \href
  {https://ui.adsabs.harvard.edu/abs/2008JGRA..113.3107P} {113, A03107}

\bibitem[\protect\citeauthoryear{{Perri} \& {Zimbardo}}{{Perri} \&
  {Zimbardo}}{2012}]{Perri12}
{Perri} S.,  {Zimbardo} G.,  2012, \mn@doi [\apj] {10.1088/0004-637X/750/2/87},
  \href {https://ui.adsabs.harvard.edu/abs/2012ApJ...750...87P} {750, 87}

\bibitem[\protect\citeauthoryear{{Perri}, {Lepreti}, {Carbone}  \&
  {Vulpiani}}{{Perri} et~al.}{2007}]{Perri07a}
{Perri} S.,  {Lepreti} F.,  {Carbone} V.,   {Vulpiani} A.,  2007, \mn@doi [EPL
  (Europhysics Letters)] {10.1209/0295-5075/78/40003}, \href
  {https://ui.adsabs.harvard.edu/abs/2007EL.....7840003P} {78, 40003}

\bibitem[\protect\citeauthoryear{{Perri}, {Greco}  \& {Zimbardo}}{{Perri}
  et~al.}{2009}]{Perri09}
{Perri} S.,  {Greco} A.,   {Zimbardo} G.,  2009, \mn@doi [\grl]
  {10.1029/2008GL036619}, \href
  {http://adsabs.harvard.edu/abs/2009GeoRL..36.4103P} {36, L04103}

\bibitem[\protect\citeauthoryear{{Perri}, {Zimbardo}  \& {Greco}}{{Perri}
  et~al.}{2011}]{Perri2011}
{Perri} S.,  {Zimbardo} G.,   {Greco} A.,  2011, \mn@doi [Journal of
  Geophysical Research (Space Physics)] {10.1029/2010JA016328}, \href
  {https://ui.adsabs.harvard.edu/abs/2011JGRA..116.5221P} {116, A05221}

\bibitem[\protect\citeauthoryear{{Perrone} et~al.,}{{Perrone}
  et~al.}{2013}]{Perrone13}
{Perrone} D.,  et~al., 2013, \mn@doi [\ssr] {10.1007/s11214-013-9966-9}, \href
  {https://ui.adsabs.harvard.edu/abs/2013SSRv..178..233P} {178, 233}

\bibitem[\protect\citeauthoryear{Petrosian}{Petrosian}{2012}]{Petrosian12}
Petrosian V.,  2012, Space science reviews, 173, 535

\bibitem[\protect\citeauthoryear{{Pisokas}, {Vlahos}, {Isliker}, {Tsiolis}  \&
  {Anastasiadis}}{{Pisokas} et~al.}{2017}]{Pisokas16}
{Pisokas} T.,  {Vlahos} L.,  {Isliker} H.,  {Tsiolis} V.,   {Anastasiadis} A.,
  2017, Astrophys. J., 835, 214

\bibitem[\protect\citeauthoryear{{Pisokas}, {Vlahos}  \& {Isliker}}{{Pisokas}
  et~al.}{2018}]{Pisokas18}
{Pisokas} T.,  {Vlahos} L.,   {Isliker} H.,  2018, \mn@doi [Astrophys. J.]
  {10.3847/1538-4357/aaa1e0}, \href
  {http://adsabs.harvard.edu/abs/2018ApJ...852...64P} {852, 64}

\bibitem[\protect\citeauthoryear{{Pontin}}{{Pontin}}{2011}]{Pontin11}
{Pontin} D.~I.,  2011, \mn@doi [Advances in Space Research]
  {10.1016/j.asr.2010.12.022}, \href
  {https://ui.adsabs.harvard.edu/abs/2011AdSpR..47.1508P} {47, 1508}

\bibitem[\protect\citeauthoryear{{Ramaty}}{{Ramaty}}{1979}]{Ramaty79}
{Ramaty} R.,  1979, in {Arons} J.,  {McKee} C.,   {Max} C.,  eds,  American
  Institute of Physics Conference Series Vol. 56, Particle Acceleration
  Mechanisms in Astrophysics. pp 135--154, \mn@doi{10.1063/1.32074}

\bibitem[\protect\citeauthoryear{{Schlickeiser}}{{Schlickeiser}}{1989}]{Schlickeiser89}
{Schlickeiser} R.,  1989, \mn@doi [\apj] {10.1086/167009}, \href
  {http://adsabs.harvard.edu/abs/1989ApJ...336..243S} {336, 243}

\bibitem[\protect\citeauthoryear{{Schlickeiser}}{{Schlickeiser}}{2003}]{Schhlickeiser02Book}
{Schlickeiser} R.,  2003, {Particle Acceleration Processes in Cosmic Plasmas}.
pp 230--260

\bibitem[\protect\citeauthoryear{{Shlesinger}, {Zaslavsky}  \&
  {Klafter}}{{Shlesinger} et~al.}{1993}]{Shlesinger93}
{Shlesinger} M.~F.,  {Zaslavsky} G.~M.,   {Klafter} J.,  1993, \mn@doi [\nat]
  {10.1038/363031a0}, \href
  {https://ui.adsabs.harvard.edu/abs/1993Natur.363...31S} {363, 31}

\bibitem[\protect\citeauthoryear{{Stawicki}}{{Stawicki}}{2005}]{Stawicki05}
{Stawicki} O.,  2005, \mn@doi [\apj] {10.1086/431662}, \href
  {https://ui.adsabs.harvard.edu/abs/2005ApJ...631..597S} {631, 597}

\bibitem[\protect\citeauthoryear{{Tsironis} \& {Vlahos}}{{Tsironis} \&
  {Vlahos}}{2005}]{Tsironis05}
{Tsironis} C.,  {Vlahos} L.,  2005, \mn@doi [Plasma Physics and Controlled
  Fusion] {10.1088/0741-3335/47/1/008}, \href
  {https://ui.adsabs.harvard.edu/abs/2005PPCF...47..131T} {47, 131}

\bibitem[\protect\citeauthoryear{{Tverskoi}}{{Tverskoi}}{1967}]{Tverskoi67}
{Tverskoi} B.~A.,  1967, Soviet Astronomy, \href
  {http://adsabs.harvard.edu/abs/1967SvA....10.1031T} {10, 1031}

\bibitem[\protect\citeauthoryear{Vlahos \& Isliker}{Vlahos \&
  Isliker}{2019}]{Vlahos18}
Vlahos L.,  Isliker H.,  2019, Plasma Physics and Controlled Fusion, 61, 014020

\bibitem[\protect\citeauthoryear{{Vlahos}, {Isliker}  \& {Lepreti}}{{Vlahos}
  et~al.}{2004}]{Vlahos04}
{Vlahos} L.,  {Isliker} H.,   {Lepreti} F.,  2004, \mn@doi [Astrophys. J.]
  {10.1086/386364}, \href {http://adsabs.harvard.edu/abs/2004ApJ...608..540V}
  {608, 540}

\bibitem[\protect\citeauthoryear{{Vlahos}, {Isliker}, {Kominis}  \&
  {Hizanidis}}{{Vlahos} et~al.}{2008}]{Vlahos08Tut}
{Vlahos} L.,  {Isliker} H.,  {Kominis} Y.,   {Hizanidis} K.,  2008, arXiv
  e-prints, \href {https://ui.adsabs.harvard.edu/abs/2008arXiv0805.0419V} {p.
  arXiv:0805.0419}

\bibitem[\protect\citeauthoryear{{Vlahos}, {Pisokas}, {Isliker}, {Tsiolis}  \&
  {Anastasiadis}}{{Vlahos} et~al.}{2016}]{Vlahos16}
{Vlahos} L.,  {Pisokas} T.,  {Isliker} H.,  {Tsiolis} V.,   {Anastasiadis} A.,
  2016, \mn@doi [Astrophys. J] {10.3847/2041-8205/827/1/L3}, \href
  {http://adsabs.harvard.edu/abs/2016ApJ...827L...3V} {827, L3}

\bibitem[\protect\citeauthoryear{{Zacharegkas}, {Isliker}  \&
  {Vlahos}}{{Zacharegkas} et~al.}{2016}]{Zacharegkas16}
{Zacharegkas} G.,  {Isliker} H.,   {Vlahos} L.,  2016, \mn@doi [Physics of
  Plasmas] {10.1063/1.4968216}, \href
  {https://ui.adsabs.harvard.edu/abs/2016PhPl...23k2119Z} {23, 112119}

\bibitem[\protect\citeauthoryear{{Zank}}{{Zank}}{2014}]{Zank14book}
{Zank} G.~P.,  2014, {Transport Processes in Space Physics and Astrophysics}.
 Vol. 877, \mn@doi{10.1007/978-1-4614-8480-6, }

\bibitem[\protect\citeauthoryear{{Zank} et~al.,}{{Zank} et~al.}{2015}]{Zank15}
{Zank} G.~P.,  et~al., 2015, \mn@doi [\apj] {10.1088/0004-637X/814/2/137},
  \href {http://adsabs.harvard.edu/abs/2015ApJ...814..137Z} {814, 137}

\bibitem[\protect\citeauthoryear{{Zhdankin}, {Uzdensky}, {Perez}  \&
  {Boldyrev}}{{Zhdankin} et~al.}{2013}]{Zhdankin13}
{Zhdankin} V.,  {Uzdensky} D.~A.,  {Perez} J.~C.,   {Boldyrev} S.,  2013,
  \mn@doi [Astrophs. J.] {10.1088/0004-637X/771/2/124}, \href
  {http://adsabs.harvard.edu/abs/2013ApJ...771..124Z} {771, 124}

\bibitem[\protect\citeauthoryear{{Zimbardo} \& {Perri}}{{Zimbardo} \&
  {Perri}}{2013}]{Zimbardo13}
{Zimbardo} G.,  {Perri} S.,  2013, \mn@doi [\apj] {10.1088/0004-637X/778/1/35},
  \href {https://ui.adsabs.harvard.edu/abs/2013ApJ...778...35Z} {778, 35}

\bibitem[\protect\citeauthoryear{{Zimbardo}, {Greco}, {Sorriso-Valvo}, {Perri},
  {V{\"o}r{\"o}s}, {Aburjania}, {Chargazia}  \& {Alexand rova}}{{Zimbardo}
  et~al.}{2010}]{Zimbardo10}
{Zimbardo} G.,  {Greco} A.,  {Sorriso-Valvo} L.,  {Perri} S.,  {V{\"o}r{\"o}s}
  Z.,  {Aburjania} G.,  {Chargazia} K.,   {Alexand rova} O.,  2010, \mn@doi
  [\ssr] {10.1007/s11214-010-9692-5}, \href
  {https://ui.adsabs.harvard.edu/abs/2010SSRv..156...89Z} {156, 89}

\bibitem[\protect\citeauthoryear{{Zimbardo} et~al.,}{{Zimbardo}
  et~al.}{2015}]{Zimbardo15}
{Zimbardo} G.,  et~al., 2015, \mn@doi [Journal of Plasma Physics]
  {10.1017/S0022377815001117}, \href
  {https://ui.adsabs.harvard.edu/abs/2015JPlPh..81f4901Z} {81, 495810601}

\makeatother
\end{thebibliography}

\end{document}